\documentclass[reprint,longbibliography,showpacs,preprintnumbers,bibnotes,amsmath,amssymb,prl]{revtex4-2}

\usepackage{graphicx}%
\usepackage[english]{babel}
\usepackage{microtype}          
\usepackage[breaklinks=true,colorlinks=true,linkcolor=blue,urlcolor=blue,citecolor=blue]{hyperref}

\usepackage{xcolor}     %
\usepackage{ulem} 
\usepackage{mathtools}
\usepackage{braket}
\usepackage{tikz}

\usepackage{longtable}
\usepackage{booktabs}
\usepackage{siunitx}
\usepackage{csquotes}
 
\usepackage{multirow}
\usepackage{mathtools}
\usepackage{amssymb}

\usepackage{dcolumn}%
\usepackage{bm}%

\begin{document}
\preprint{APS/123-QED}

\title{Ultrafast Charge Separation  Induced by a Uniform Field in Graphene Nanoribbons}

\author{Jan-Philip Joost, and
Michael Bonitz
 \email{joost@theo-physik.uni-kiel.de}}
\affiliation{
Institut f\"ur Theoretische Physik und Astrophysik, 
Christian-Albrechts-Universit\"{a}t zu Kiel, D-24098 Kiel, Germany
}

\date{\today}%

\begin{abstract}
When heteronuclear molecules are illuminated by light of spatially uniform intensity, electronic excitations may, nevertheless, be restricted to parts of the system, depending on the absorption properties of its constituents.
Here, we show that this effect is observed also in homogenous carbon based systems, such as graphene nanoribbons (GNRs):
 a spatially uniform laser pulse can create strongly localized carrier excitations, including excitons, on the sub-nanometer scale within a few femtoseconds. The origin of this effect is the unusual topological-based electronic structure of the GNRs.
 This opens new avenues for nanoelectronics and brings petahertz switching within reach. Using nonequilibrium Green functions simulations we demonstrate this effect by exciting small GNR heterostructures of suitable geometry with a laser pulse with carefully chosen photon energy,  polarization, and carrier-envelope phase.
\end{abstract}

\maketitle
%
%
%
Progress in nanotechnology is currently severely limited by electronic switching speeds in the GHz range. To overcome this,  various ideas have been put forward that use single-cycle optical pulses which would achieve petahertz switching. Rybka \textit{et al.} demonstrated coherent light wave control of electronic currents in plasmonic nanocircuits~\cite{rybka_nphot_16}. This was extended by Keathley \textit{et al.} to photoemission from gold nanoantennas \cite{keathley_np_19}. Light field driven real and virtual carriers were reported by Hommelhoff and co-workers in Ref.~\cite{boolakee_light-field_2022}, who also demonstrated the important role of electronic correlation effects in ultrafast photoemission 
\cite{meier_np_23}.
Subfemtosecond light-driven charge dynamics  were reported in Refs.~\cite{Pettine2024} and \cite{Inzani2023}.
A second direction of progress is to exploit the potential of novel quantum materials of reduced dimensionality, such as monolayers of graphene or transition metal dichalcogenides (TMDCs). They offer a very broad range of electronic and optical properties including strong excitonic effects
\cite{chernikov_prl_14,moody_ncom_15,na_sc-20,wang_rmp_18,Jiang2023}.
Additional degrees of freedom emerge when the system dimensions are further reduced to only a few nanometers and the shape of the monolayer clusters is varied.
Particularly promising candidates are nanoclusters of graphene or TMDCs, graphene nanoribbons (GNRs) \cite{fujita_jpsj_96,son_energy_2006,ElAbbassi2020,Wang2021}, and nanographenes \cite{gu_jacs_22}. 

Here we concentrate on GNRs because of their interesting electronic properties \cite{son_energy_2006}, which are controlled by correlation effects \cite{prezzi_optical_2008, joost_19_nanolett}, 
as well as strong excitonic effects \cite{yang_excitonic_2007}.
Zigzag GNRs were shown to host ferromagnetically ordered edge states \cite{blackwell_nature_2021,song_nature_25}, whereas topological junction states in GNRs have been proposed as molecule sensing devices \cite{abdelsalam_xxx_24}.
Of particular interest is the appearance of topological in gap states that are spatially localized at particular points of the GNRs \cite{rizzo_topological_2018,joost_19_nanolett,Slicker2024}. 
This non-trivial electronic structure is expected to give rise to interesting response behavior, when GNRs are excited by a laser pulse. Indeed, in Ref.~\cite{boolakee_light-field_2022} it was demonstrated that electron generation can be controlled by the carrier-envelope phase (CEP) of the laser, whereas 
 Ref.~\cite{wang_prb_24} reported excitation of large electrical currents along the GNR axis for special pulse forms.
In this Letter we predict and demonstrate a different effect: excitation of spatially localized electrons or excitons, on the time scale of a few femtoseconds, the location of which can be controlled by the laser frequency, polarization, and carrier-envelope phase. This effect occurs even when the GNR heterostructure is excited with a spatially uniform laser pulse. Creation of charge separation and currents by uniform laser pulses is well known in heteronuclear molecules such as light-harvesting complexes \cite{fassioli_jrsi_14} or nitrogen vacancy centers in diamond~\cite{DOHERTY_pr_13},  where the effect arises from the different electronic properties of the respective constituents. However, in contrast to these systems, here we consider GNRs which are structurally homogeneous and where charge localization is a consequence of the unusual topology-based electronic structure~\cite{rizzo_topological_2018,joost_19_nanolett}. 

%
\begin{figure*}[ht]
\includegraphics[width=\textwidth]{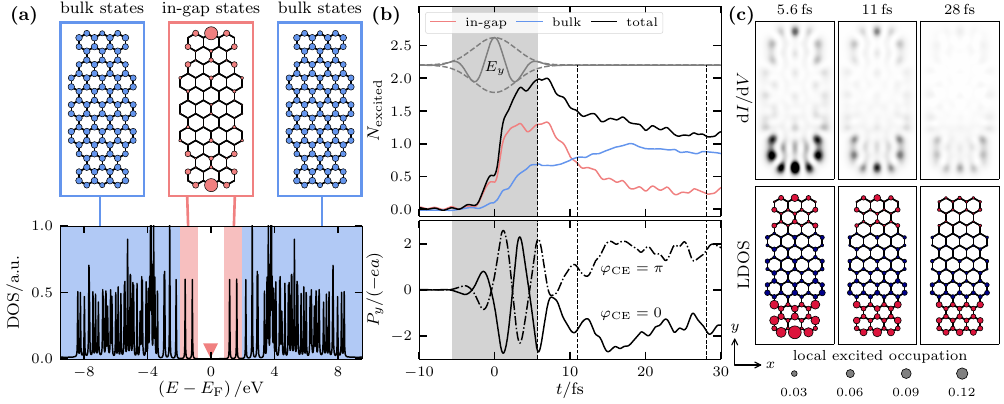}
  \caption{\label{fig:1}(a) Density of states (bottom) and spatial localization (top) of the indicated states of System 1. (b) Time-dependent carrier density in the bulk (blue) and edge region (red) and their sum (black) in response to a short circularly polarized laser pulse (top) and $y$-component of the dipole moment (bottom) for two values of the CEP. (c) Space resolved differential conductance at $4\,$\AA$\,$ above the GNR \cite{joost_19_nanolett} (top) and occupied LDOS, for three times (bottom). 
  The top-bottom asymmetry is controlled by the carrier-envelope phase: $\phi_{\rm CE}=0$ corresponds to the $E_y$ shown in (b), whereas for $\phi_{\rm CE}=\pi$, $E_y$ has its minimum at $t=0$. 
  }
\end{figure*}

\begin{figure}
\includegraphics[width=\columnwidth]{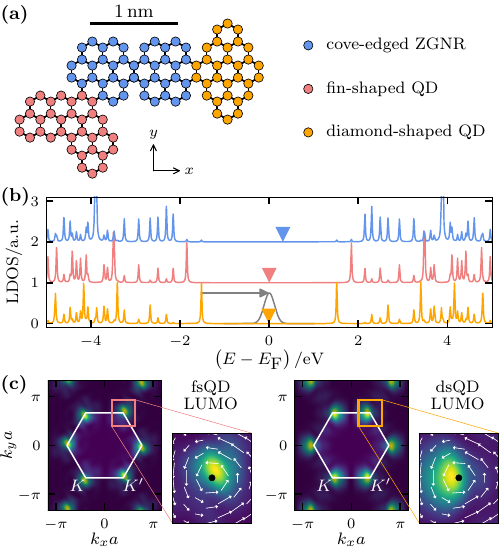}
  \caption{\label{fig:2}Ground state properties of System 2 (a): a short cove-edged ZGNR (blue) terminated by two different graphene quantum dots (red and yellow). (b) LDOS of the three system parts of (a) (the same colors). 
The respective excitonic states are marked by triangles. 
(c) Momentum distribution of the LUMO states of the two quantum dots. The first Brillouin zone is indicated by the white hexagon. The white arrows represent the interband optical matrix element $\bm{M}^{v,c}_{\bm{k}}$ in dipole approximation for tight-binding pristine graphene~\cite{bonitz_pssb23}.
}
\end{figure}

\begin{figure}
\includegraphics[width=\columnwidth]{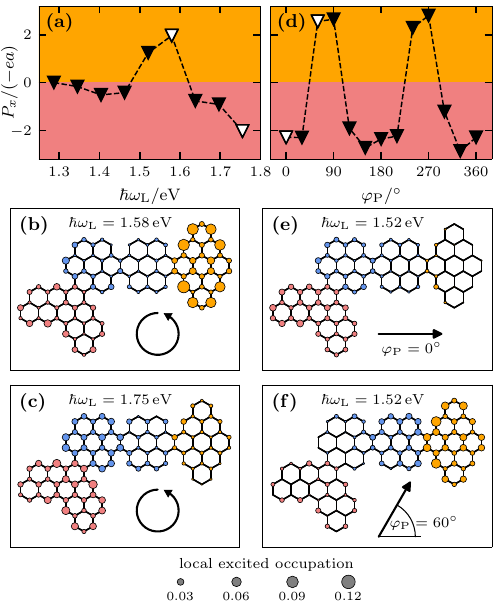}
  \caption{\label{fig:3} Dependence of spatial charge separation in System 2 [measured by the horizontal component of the dipole moment], on (a) photon energy, for  circular polarization, and  (d) laser polarization, for $\hbar \omega_{\textnormal{L}}=1.52\,$eV, at $t=28\,\mathrm{fs}$, i.e., long after the laser pulse. (b, c, e, f): Spatial distribution of the excited carriers at $t=28\,\mathrm{fs}$ 
  for four combinations of polarization and $\omega_{\rm L}$ indicated in the figures [white triangles in (a) and (d)]. 
}
\end{figure}

\textit{Model}---We model the graphene nanostructures subject to a uniform laser pulse with field strength $\bm{\mathcal{E}}(t)$
by describing the $\pi$ electrons within the Pariser--Parr--Pople (PPP) lattice model~\cite{chiappe_can_2015} which includes an on-site energy term ($\epsilon$), nearest neighbor hopping with amplitude $J$, Hubbard-type onsite interaction ($U$) and long-range Coulomb interactions ($V_{ij}$),
\begin{align}
 &     \hat{H}(t) =
      \epsilon \sum_{i,\sigma} \hat{c}^\dagger_{i,\sigma} \hat{c}_{i,\sigma}  - J \sum_{\langle i,j\rangle,\sigma} \hat{c}^\dagger_{i,\sigma} \hat{c}_{j,\sigma}
      + U \sum_{i} \hat{n}_{i,\uparrow} \hat{n}_{i,\downarrow} +
       \nonumber\\
      & \frac{1}{2} \sum_{i \neq j,\sigma,\sigma^\prime} V_{ij} \left(\hat{n}_{i,\sigma}-1\right) \left(\hat{n}_{j,\sigma^\prime}-1\right)
      + e \sum_i \bm{r}_i \cdot \bm{\mathcal{E}}(t)\,\hat{n}_i,
    \label{eq:h}
\end{align}
where the latter are interpolated using the Ohno parametrization~\cite{ohno_remarks_1964}. The operators $\hat{c}^\dagger_{i,\sigma}$ and $\hat{c}_{i,\sigma}$ create and annihilate an electron of spin $\sigma$ at lattice site $i$, and $\hat{n}$ is the particle number operator. The set of model parameters is taken from Ref.~\cite{verges_fit_2010} as $J=2.34\,\mathrm{eV}$, $\epsilon=-3.25\,J$ and $U=3.54\,J$, which was shown to closely reproduce the \textit{ab initio} $GW$ band gaps of a wide range of freestanding graphene nanoribbons~ \cite{joost_phd_2022}. 
Screening of the local and long-range interactions by an underlying surface is taken into account by rescaling the interaction parameter $U$. For Figs.~\ref{fig:1}--\ref{fig:3} this effective interaction is set to a significantly reduced value of $U/J=2$, which corresponds to strong external screening by a metallic substrate, such as Au(111)~\cite{joost_femtosecond_2019,joost_19_nanolett}. More information on the PPP model and the incorporation of surface screening can be found in the SM~\cite{supplemental_material}.
The electron dynamics of system \eqref{eq:h} that is induced by a spatially uniform short laser pulse is computed using the recently developed G1--G2 scheme, an efficient time-linear Nonequilibrium Green functions (NEGF) approach \cite{schluenzen_prl_20,joost_prb_20,bonitz_pssb23}. We employ the DSL approximation \cite{joost_prb_22} that accounts for dynamical screening and strong correlation effects simultaneously ($GW$ plus vertex corrections) and applies to conditions far from equilibrium. The time-dependent spectral information is obtained by using Koopmans' theorem \cite{koopmans_uber_1934,niggas_ion-induced_2022,joost_phd_2022}. Additional information on the G1--G2 scheme is given in the SM~\cite{supplemental_material}.

\textit{7--9-AGNR unit cell (System 1)}---The first system that we consider (``System 1'', Fig.~\ref{fig:1}) is the unit cell of the 7--9-AGNR heterostructure,  the ground-state properties of which were studied in depth both, experimentally and theoretically \cite{rizzo_topological_2018,joost_19_nanolett,rizzo_rationally_2021,louis_long_2024}. The strongly localized topological in-gap edge states make this system a prime candidate for a localized excitation of charge carriers, even with a spatially extended optical pulse. In Fig.~\ref{fig:1}a the DOS of system 1 is shown and the spatial extension of both,  edge states (localized on the outer zigzag edges, red) and bulk states (extended across the system, blue), is highlighted. 
 The energy of the bound exciton state with binding energy $E_\mathrm{b} =  1.2\,\mathrm{eV}$ is marked in Fig.~\ref{fig:1}a by the triangle inside the band gap.

In Fig.~\ref{fig:1}b we present simulation results for an excitation of System 1 that specifically targets the in-gap edge states. We use a circularly polarized, 
$1.5$-cycle
laser pulse (top figure) with amplitude $\mathcal{E}_0 = 1.65\,\mathrm{V/nm}$ [fluence $1.8\,\mathrm{mJ/cm}^2$], pulse duration (FWHM) of $6.6\,\mathrm{fs}$
and photon energy $\hbar\omega = 0.7\,\mathrm{eV}$ 
which ensures that only edge states are excited.
As shown in Fig.~\ref{fig:1}b, indeed the edge state population rises quickly and reaches its maximum already shortly after the peak of the laser pulse. During a second phase, bulk states are excited, driven by the large population of edge states and edge-bulk state overlap,
so that after the laser pulse, at a time of $5.6\,\mathrm{fs}$, the bulk state occupation reaches half the value of the edge states.
The subsequent dynamics are characterized by the relaxation of the system into a pre-thermal state mediated by Auger-like scattering events. During this process, additional bulk electrons are excited, while the number of excited edge electrons rapidly declines. Already at $11\,\mathrm{fs}$ both edge and bulk states have comparable occupations and at $28\,\mathrm{fs}$ only $20\,\%$ of the initially excited edge state electrons remain.

Interestingly, despite the top-bottom symmetry of System 1, the laser pulse predominantly excites edge states in the bottom part, cf. Fig.~\ref{fig:1}c. At $t=5.6\,\mathrm{fs}$ the average occupation per site,
    $\rho^{\alpha}_{\rm excited}(t) = N^{\alpha}_{\rm excited}(t) / N^{\alpha}_{\rm sites}$ for region $\alpha\,,$
is $0.06$ in the bottom red region, compared to $0.02$ in the upper one. Despite the aforementioned short lifetime of the edge states, this asymmetry persists even at $t=28\,\mathrm{fs}$ where the average occupation is $0.03$ at the bottom and $0.01$ at the top. This is explained by the chosen carrier envelope phase (CEP) of the laser, $\phi_{\rm CE}=0$, which corresponds to $\mathcal{E}_y$ being maximal at $t=0$, cf. the top part of Fig.~\ref{fig:1}b. The influence of the CEP is demonstrated in the bottom panel of Fig.~\ref{fig:1}b, where the y-component of the dipole moment of the excited electrons,
$\bm{P}(t) = \sum_{i=1}^{N_{\mathrm{sites}}} q_i(t) \bm{r}_i$,
is shown. After initial oscillations, the dipole moment at $t=11\,\mathrm{fs}$ reaches and retains a value of about $-2$, in units of the electron charge and lattice constant $a$. The effect can be reversed by using a laser pulse with a field minimum at the peak intensity, as demonstrated by the results for the dipole moment for $\phi_{\rm CE}=\pi$ (dash-dotted line in Fig.~\ref{fig:1}b). A similar effect was observed for charge-polarization originating from virtual charge carriers in macroscopic graphene nanoribbons connected to metallic contacts~\cite{boolakee_light-field_2022} and the utilization of the CEP for symmetry breaking has been discussed in detail in Refs.~\cite{franco_jpb_08,schiffrin_nature_13}
%

To summarize, 
we demonstrated that a spatially uniform short laser pulse that targets topological in-gap states, indeed allows for a localized excitation of electrons in the edge part of System 1, within about $5\,\mathrm{fs}$. However, this bulk-edge charge separation quickly decays thereafter. In addition we observed a second effect which is determined by the carrier-envolope phase of the laser: a stable top-down charge separation within the edge regions that gives rise to a large dipole moment component, $P_y$.

\textit{Quantum dot heterostructure (System 2)}---We now inquire whether the bulk-edge charge separation can be stabilized for longer times. To this end, we design a different geometry of the graphene cluster (System 2),
that is displayed in Fig.~\ref{fig:2}a. It consists of a short central region, reminiscent of a cove-edged ZGNR~\cite{lee_topological_2018,arnold_structure-imposed_2022} (blue), and two terminating quantum dots (QD) of different geometries, which we denote as fin-shaped (red) and diamond-shaped (yellow). Graphene nanostructures of comparable size can be reliably synthesized with atomic precision~ \cite{mishra_topological_2020,mishra_observation_2021,sun_molecular_2022,barin_on-surface_2023}.
In Fig.~\ref{fig:2}b we display the LDOS which  
confirms that there is little overlap between the three individual local spectra. 
This is an effect of both the unique geometries of the QDs and the cove-edge structure of the system, which creates bottlenecks in the lattice between the different regions. 
In addition, in Fig.~\ref{fig:2}b the energy of the lowest bound exciton state is marked by a triangle of the respective color, for each part of the system, corresponding to a binding energy, $E_\mathrm{b} = 1.52\,\mathrm{eV}, 1.84\,\mathrm{eV}, 1.84\,\mathrm{eV}$, for the yellow, red and blue parts, respectively. The gray envelope illustrates the spectral width of the laser used for the calculations of Fig.~\ref{fig:3}.
Our strategy to achieve spatially localized charge distributions is to selectively excite electrons into these excitonic states, which can be realized by a specific choice of not only the laser energy but also its polarization. The latter is explained by the distinct orientation of the QDs, which are connected to the central region at different angles. 
Since the interband optimal matrix elements of the system $\bm{M}^{v,c}_{\bm{k}}$, displayed in Fig.~\ref{fig:2}c, couple to the laser vector potential via $\bm{M}^{v,c}_{\bm{k}}\cdot \bm{A}(t)$, the optimal polarization angle for optical excitation at a given $\bm{k}$ is determined by the orientation of $\bm{M}^{v,c}_{\bm{k}}$. Therefore, electrons are expected to be primarily excited within the yellow (red) QD for $\varphi_\mathrm{P} = 60^\circ$ to $90^\circ$ and $\varphi_\mathrm{P} = 240^\circ$ to $270^\circ$ ($\varphi_\mathrm{P} = 150^\circ$ to $180^\circ$ and $\varphi_\mathrm{P} = 330^\circ$ to $360^\circ$).
Details on the $\bm{M}^{v,c}_{\bm{k}}$ are presented in the SM~\cite{supplemental_material}.

In Fig.~\ref{fig:3} we demonstrate that it is, indeed, possible to excite electrons selectively in the red or yellow QD, by tuning the frequency (a-c) and polarization (d-f) of a short laser pulse,
%
and that the resulting charge separation remains stable
long after the laser pulse ($t=28\,$fs).
As shown in Fig.~\ref{fig:3}a by means of the x-component of the dipole moment, for circular polarization choosing a photon energy in the range of $\hbar \omega_\mathrm{L}=1.5\,$eV to $1.6\,$eV excites predominantly carriers in the yellow QD (cf. Fig.~\ref{fig:3}b), whereas photons with $\hbar \omega_\mathrm{L}=1.65\,$eV to $1.8\,$eV excite predominantly charges in the red QD [cf. Fig.~\ref{fig:3}c] and, due to the spectral width of the laser, also carriers in the blue region.
For the choice of linear polarization, the results are in good agreement with the theoretical predictions derived from Fig.~\ref{fig:2}c. For angles from 45 to 105 and 225 to 285 degrees, electrons are excited in the yellow QD and, in all other cases, in the red QD. Two examples for $\varphi_\mathrm{P}=0^\circ$ and $\varphi_\mathrm{P}=60^\circ$ are shown in Figs.~\ref{fig:3}e and f, respectively. 
%
In all studied cases, charge separation is a direct consequence of the electron--laser interaction and is not significantly modified by the subsequent scattering dynamics which is characteristic for strongly screened systems. In the following, we study an unscreened, free-standing system.


\begin{figure}[t]
\includegraphics[width=\columnwidth]{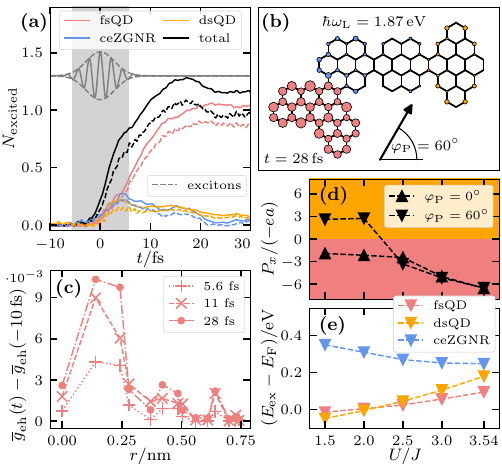}
  \caption{\label{fig:5}Excitation of 
System 2 (free-standing, $U/J=3.54$)
  with a four-cycle laser pulse of frequency $\hbar\omega_{\textnormal{L}}=1.87\,$eV with linear polarization [arrow in (b)]. (a) Laser pulse and time evolution of the number of excited carriers (full lines) and excitons  (dashed lines) in the three system parts. 
  (b) Spatial distribution of the excited electrons at $t=28\,$fs. 
  (c) Average e-h pair correlation function, $\bar g_{eh}$, in the red QD  for three times. (d) Dipole moment for and two linear polarizations vs. Hubbard $U$. (e) $U$-dependence of the exciton energy in the three system parts, cf. Fig.~\ref{fig:2}b.  
  In panels (d) and (e) the laser frequency changes from $\hbar\omega_{\textnormal{L}}=1.4\,$eV for $U/J=1.5$, to $\hbar\omega_{\textnormal{L}}=1.9\,$eV for $U/J=3.54$, for details see text.}
\end{figure}


\textit{Free-standing system}---We analyze the time evolution of the charge separation for the case $U/J=3.54$
which is presented in Fig.~\ref{fig:5}a. The laser parameters are the same as for Fig.~\ref{fig:3}f with $\varphi_\mathrm{P}=60^\circ$, but with $\hbar\omega_{\textnormal{L}}=1.87\,$eV, to account for the modified band structure of the free-standing system. Over the first $5\,$fs, carriers are excited in all three system parts. However, after this initial phase the carrier number in the blue and yellow segments decreases, whereas, in the red QD, it continues to increase, even after the pulse, reaching a stable maximum around $20\,$fs. 
Thus, not only does the laser pulse generate a pronounced charge separation already within only a few femtoseconds, but this charge separation further increases and remains remarkably stable, changing only slightly over a period of $30\,\mathrm{fs}$. However, the final charge distribution [Fig.~\ref{fig:5}b] differs significantly from that for the screened setup, cf.~Fig.~\ref{fig:3}f, where predominantly the yellow QD gets excited.

Let us, therefore, now explore the physical properties of the locally accumulated charge. Since, in Fig.~\ref{fig:5}, the laser energy is being pumped directly into an exciton state (analogous to Fig.~\ref{fig:3}, cf. Fig.~\ref{fig:2}b) one could expect that predominantly excitons are being formed.  
To verify this hypothesis we compute the 
locally averaged electron--hole pair correlation function (PCF), $\bar g_{eh}(r)$\footnote{The electron--hole PCF is defined as $\bar{g}_\mathrm{eh}^\alpha(r) = \frac{1}{N^\alpha_\mathrm{sites}} \sum_{R\in \alpha} g_\mathrm{eh}(r;R)$, where $r$ and $R$ are, respectively, relative and center of mass coordinates of the pair, and the sum (averaging) runs over the system part $\alpha$.}.
The laser-induced change of $\bar g_{eh}(r)$, for the red QD, is shown in Fig.~\ref{fig:5}b for three different times. 
The maximum of $\bar g_{eh}(r)$ forms between $0.14$ and $0.25\,$nm with its amplitude steadily increasing, up to $t=28\,$fs, indicating that the excitons are tightly bound within a radius of approximately one lattice constant. The integral over $\bar g_{eh}(r)$ provides a measure for the number of created excitons, cf. dashed lines in Fig.~\ref{fig:5}a. During both the laser interaction and the subsequent scattering dynamics excitons make up the majority of the excited charges. 
%
%
Thus, we conclude that the presented excitation scenario creates spatially localized excitons in the red quantum dot that remain stable for at least $30\,$fs.

The dependence of the charge separation on  $U$ (i.e., on substrate screening) is further explored in Fig.~\ref{fig:5}d by means of the induced dipole moment. The laser frequency is chosen as $\hbar\omega_{\textnormal{L}}=1.4\,$eV for $U/J=1.5$ and increased by $0.12\,$eV for each $\Delta U/J=0.5$ to account for the shift of the electronic levels, cf.~Fig.~\ref{fig:5}e. While, for $U/J \leq 2$, the laser polarization can be used to selectively excite charges in the yellow or red QD (cf.~Fig.~\ref{fig:3}d), for large $U$,  the ultrafast electron dynamics always result in charges accumulating in the red QD, with the transition occurring for $ 2\le U/J \le 2.5$.
While in strongly screened systems e--e scattering between different system parts is reduced, preventing the redistribution of charges across the system, 
for free-standing systems the strong long-range Coulomb interactions allow for efficient charge transfer between different regions.
%
Since for the corresponding interaction strengths ($U/J\geq 2.5$) the bound exciton state of the red QD has the lowest energy, cf.~Fig.~\ref{fig:5}e, most excited charges accumulate there, through Auger-like scattering.



\textit{Summary and discussion}---
To summarize, in this Letter we have demonstrated that a spatially uniform laser pulse can create stable spatial charge separation on the sub-nanometer scale within a few femtoseconds 
in graphene nanoribbons -- an effect that was previously observed only in heterogeneous systems, such as NV-centers in diamond~\cite{DOHERTY_pr_13} or light-harvesting complexes~\cite{fassioli_jrsi_14}.
The mechanism behind this effect 
in GNRs
is (1) the existence of a pronounced space dependence of the DOS due to isolated topological states, (2) a proper choice of laser energy and polarization that allows for excitation of those states and (3) measures like spatial bottlenecks of surface screening to reduce to stabilize the charge separation in the homogeneous system. We presented direct evidence for this effect by performing extensive NEGF simulations that take into account strong coupling and dynamical screening. 
While for a cluster of simple geometry (System 1) charge separation between bulk and edge regions is achieved quickly (within $5\,$fs), but is lost within the next $5\,$fs, we presented a second cluster (System 2) where the effect persists for at least $30\,$fs and can be controlled,  to a large extent, by proper choice of the laser parameters. Interestingly, the dominant fraction of localized electrons are bound in excitons. An additional control for charge separation in System 1 was found to be the choice of the carrier-envelope phase.

Our predictions are not limited to the two chosen systems. Similar behavior is expected for a broad class of nanoscale monolayer clusters of graphene or TMDC if a number of requirements are met. First, 
 in order to create stable charge separation, the local states should be highly confined by spatial bottlenecks (as in the case of System 2) to minimize orbital overlap with other regions of the system. 
Second, surface screening plays an important role. While the simple picture described above holds for screened systems  ($U/J \lesssim 2$), for large $U$, such as 
for free-standing systems, charge separation is still possible,
but is strongly affected by e--e interactions. This impairs the direct selection of the areas to be excited by choice of the laser parameters, cf. Fig.~\ref{fig:5}.
Third, for the use of linearly polarized light it is required to fix the spatial orientation of the clusters on the substrate. Finally, the maximum duration of the charge separation, beyond the demonstrated $30\,$fs, will be limited by the coupling to phonons. 
The latter are straightforwardly included in the G1--G2 scheme, see e.g., Refs.~\cite{pavlyukh_time-linear_2022-1,pavlyukh_time-linear_2022,pavlyukh_pss_23}, for more details, see the supplemental material~\cite{supplemental_material}.


We expect that the results of our paper will open new avenues for nanoelectronics, bringing switching with petahertz frequency within reach. In contrast to existing concepts for PHz electronics, where typically symmetric systems are used and current switching is achieved via CEP control \cite{franco_jpb_08,schiffrin_nature_13,boolakee_light-field_2022}, the present GNR structures offer additional flexibility: charge separation is additionally controlled via photon energy and laser polarization, as well as optimization of the GNR geometry and asymmetry (System 2). Moreover, the small system size of the GNR allows one to use comparatively low laser energies.
Thereby the applications proposed in this Letter benefit from the topological character of the in-gap states that provides protection against noise and thermal effects. Moreover, owing to their chemically homogeneous structure, GNRs can be synthesized with atomic precision enabling the design of nanostructures with selected electronic and topological properties~\cite{yao_synthesis_2021,wenhui_bottom-up_2023,liu_cove-edged_2024,song_nature_25}.
%

There are practical challenges to be addressed: novel ideas may be required to  either contacting the localized charge or to access the dipole moment of the system.
Moreover, the excitonic nature of the localized charge allows for optical coupling and information storage, whereas the bosonic nature or the excitons  has the potential to exploit quantum coherence effects and superfluidity \cite{filinov_prl_10, pereira_nphys-22, cutshall_sci-adv_25}.



Acknowledgements.
We thank F.~Caruso for valuable comments. This work was supported by the Deutsche Forschungsgemeinschaft via grant BO1366/16.

\bibliography{library,mb-ref}

\end{document}